\documentclass[a4paper,11pt]{article}
\usepackage{pos}
\usepackage{subcaption}

\newcommand\as{\alpha_{\mathrm{S}}} 

\title{Mixed QCD--EW corrections to Drell-Yan at the LHC}
\author*[a]{Luca Buonocore}

\affiliation[a]{Universit\"at Z\"urich, Physik Institut\\
  Winterthurerstrasse 190, CH-8057 Z\"urich, Switzerland}

\emailAdd{lbuono@physik.uzh.ch}

\abstract{We discuss recent theoretical results for the complete mixed
  QCD--EW corrections to lepton-pair production via the Drell-Yan
  mechanism, a cornerstone process for the precision physics programme
  at the LHC.  We present results for fiducial cross sections and
  differential distributions to both the neutral current- and charged
  current- process. We show that the mixed corrections are relatively
  large and of similar size to the NNLO QCD ones at central scales. In
  particular, for the neutral current case, we report the first result
  at this order that is valid in the entire range of dilepton
  invariant masses.}

\FullConference{%
  *** The European Physical Society Conference on High Energy Physics (EPS-HEP2021), ***\\
  *** 26-30 July 2021 ***\\
  *** Online conference, jointly organized by Universität Hamburg and the research center DESY ***
}

\begin{document}
\maketitle

\section{Introduction}

In view of the absence of striking signals for beyond the Standard
Model phenomena at the CERN Large Hadron Collider (LHC), alternative
routes must be pursued to uncover possible new physics effects. A
viable strategy is the search for small deviations from the
predictions of the Standard Model and precision is the key for this
path.

Lepton-pair production through an off-shell vector boson, namely
the Drell-Yan mechanism, represents a cornerstone process for the
precision physics programme at the LHC. It features large production
rates and clean experimental signatures, given the presence of at
least one lepton with large transverse momentum in the final
state. Therefore, it is of primary importance for the extraction of
crucial electro-weak parameters of the Standard Model, as the $W$
boson mass and the Weinberg angle, as well as input to the
determination of parton distribution functions (PDFs).  Furthermore,
the Drell-Yan process is also very important in the context of new
physics searches, severely constraining many possible scenarios.

The Drell-Yan process is one of the most studied and well known processes.
Inclusive cross sections have been recently pushed to the N$^3$LO
level for the production of a virtual
photon~{\cite{Duhr:2020seh,Chen:2021vtu}} and of $W$
boson~\cite{Duhr:2020sdp}. Since predictions for differential
observables including leptonic decays are known up to NNLO in
QCD %~\cite{Anastasiou:2003yy,Anastasiou:2003ds,Melnikov:2006kv,Catani:2009sm,Catani:2010en}
 and and up to NLO in
EW, %~\cite{Dittmaier:2001ay,Baur:2004ig,Zykunov:2006yb,Arbuzov:2005dd,CarloniCalame:2006zq,Baur:2001ze,Zykunov:2005tc,CarloniCalame:2007cd,Arbuzov:2007db,Dittmaier:2009cr}
%and the high-precision determination of EW parameters requires
%control over the kinematical distributions at very high accuracy,
the attention of the theory community has recently turned to the mixed
QCD--EW corrections.

In this contribution, we will present recent results for the mixed corrections to
both neutral current- and charged current-  Drell-Yan process, showing their impact
on fiducial cross sections and on a selection of kinematical distributions. 

\section{Neutral current Drell-Yan process}

We focus on the hadroproduction of a massive lepton-anti-lepton pair, namely the process
\begin{equation}\label{eq:proc}
  pp\to \ell^+\ell^-+X.
\end{equation}
and we introduce the following notation for the corresponding differential cross section
\begin{equation}
  \label{eq:exp}
  d{\sigma}=\sum_{m,n=0}^\infty d{\sigma}^{(m,n)}\, ,
\end{equation}
where $d{\sigma}^{(0,0)}\equiv d{\sigma}_{\rm LO}$ is the Born level
contribution and $d{\sigma}^{(m,n)}$ the ${\cal O}(\as^m\alpha^n)$
correction.  In particular, the term $m=n=1$ corresponds to mixed
QCD--EW corrections.

The computation of ${\cal O}(\as\alpha)$ mixed corrections to the full
$2\to2$ Drell-Yan process Eq.~\eqref{eq:proc} is a formidable
task. The evaluation of the corresponding $2\to 2$ two-loop amplitudes
involves scalar integrals with internal masses at the frontier of
current computational techniques. Furthermore, it presents a rich
infra-red (IR) structure involving both initial- and final- state
radiation. The cancellation of the corresponding IR singularities
requires a suitable NNLO subtraction scheme.

Very recently the exact mixed corrections have been achieved in
Ref.~\cite{Bonciani:2021zzf}. The two-loop virtual contribution has
been worked out by using semi-analytical techniques, overcoming the
technical problems in the evaluation of the relevant master
integrals. The cancellation of the IR divergences has been achieved by
exploiting the $q_T$ subtraction formalism~\cite{Catani:2007vq}. The
extension of the method to the NLO EW and the mixed QCD--EW
corrections have been worked out in Refs.~\cite{Buonocore:2019puv}
and~\cite{Buonocore:2021rxx}, starting from the $q_T$ subtraction
formalism for heavy quarks~\cite{Catani:2019iny}.

\subsection{Results}
We show results for the process $pp\to \mu^+\mu^-+X$ at centre-of-mass
energy \mbox{$\sqrt{s}=14$\,TeV} with the setup of
Ref.~\cite{Bonciani:2021zzf}. In particular, we use the following
selection cuts,
\begin{equation}
  \label{eq:cuts}
p_{T,\mu^\pm}>25\,{\rm GeV}\,,\quad |y_{\mu^\pm}|<2.5\,,\quad
m_{\mu\mu}>50\,{\rm GeV}\, .
\end{equation}
We work at the level of {\it bare muons}, i.e., no lepton
recombination with close-by photons is carried out. In
Table~\ref{tab:fid-nc} we present predictions for the corresponding
fiducial cross section at central scales $\mu_{\rm R}=\mu_{\rm F}=m_Z$. We show the breakdown of the different
contributions $\sigma^{(i,j)}$ into the various partonic channels.
The contribution from quark--antiquark annihilation is denoted by
$q{\bar q}$.
%%====================================
\begin{table}
\renewcommand{\arraystretch}{1.1}
\centering
\begin{subtable}[b]{0.475\textwidth}
  \resizebox{0.995\textwidth}{!}{
  \begin{tabular}[b]{|c|c|c|c|c|c|}
    \hline
    $\sigma$ [pb] & $\sigma_{\rm LO}$  & $\sigma^{(1,0)}$  & $\sigma^{(0,1)}$  & $\sigma^{(2,0)}$ & $\sigma^{(1,1)}$ \\
    \hline
    \hline
    $q{\bar q}$ & $809.56(1)$ & $191.85(1)$ & $-33.76(1)$ & $49.9(7)$ & $-4.8(3)$\\
    \hline
    $qg$ & --- & $-158.08(2)$ & --- & $-74.8(5)$ & $8.6(1)$ \\
    \hline
    $q(g)\gamma$ & --- & --- & $-0.839(2)$ & --- & $0.084(3)$\\
    \hline
    $q({\bar q})q^\prime$ & --- & --- & --- & $6.3(1)$ & $0.19(0)$\\
    \hline
    $g g$ & --- & --- & --- & $18.1(2)$ & --- \\
     \hline
    $\gamma \gamma$ & $1.42(0)$ & --- & $-0.0117(4)$ & --- & --- \\
    \hline
    \hline
    tot & $810.98(1)$ & $33.77(2)$ & $-34.61(1)$ & $-0.5(9)$ &  $4.0(3)$ \\
    \hline
    \hline
    $\sigma/\sigma_{\rm LO}$ & $1$ & $4.2\%$ &  $-4.3\%$ &  $~0\%$  &   $+0.5\%$  \\
    \hline
  \end{tabular}
  }
  \caption{\label{tab:fid-nc} $pp\to\mu^+\mu^-+X$.}
\end{subtable}
\begin{subtable}[b]{0.515\textwidth}
  \resizebox{\textwidth}{!}{
    \begin{tabular}[b]{|c|c|c|c|c|c|}
    \hline
    $\sigma$ [pb] & $\sigma_{\rm LO}$  & $\sigma^{(1,0)}$  & $\sigma^{(0,1)}$  & $\sigma^{(2,0)}$ & $\sigma^{(1,1)}$ \\
    \hline
    \hline
    $q{\bar q}$ & $5029.2$ & $\phantom{+0}970.5(3)\phantom{00}$ & $-143.61(15)$ & $\phantom{+}251(4)\phantom{.0}$ & $\hspace*{-0.6ex}\phantom{0}-7.0(1.2)\phantom{00}\hspace*{-0.6ex}$\\
    \hline
    $qg$ & --- & $-1079.86(12)$ & --- & $-377(3)\phantom{.0}$ & $\hspace*{-0.6ex}\phantom{+}39.0(4)\phantom{.00}\hspace*{-0.6ex}$ \\
    \hline
    $q(g)\gamma$ & --- & --- & $\phantom{+00}2.823(1)$ & --- & $\hspace*{-0.6ex}\phantom{+0}0.055(5)\phantom{.0}\hspace*{-0.6ex}$\\
    \hline
    $q({\bar q})q^\prime$ & --- & --- & --- & $\phantom{+0}44.2(7)$ & $\hspace*{-0.6ex}\phantom{+0}1.2382(3)\phantom{.}\hspace*{-0.6ex}$\\
    \hline
    $g g$ & --- & --- & --- & $\phantom{+}100.8(8)$ & --- \\
    \hline
    \hline
    tot & $5029.2$ & $\phantom{0}$$-109.4(4)\phantom{00}$ & $-140.8(2)\phantom{00}$ & $\phantom{00}19(5)\phantom{.0}$ &  $\hspace*{-0.6ex}\phantom{+} 33.3(1.3)\phantom{00}\hspace*{-0.6ex}$ \\
    \hline
    \hline
    $\sigma/\sigma_{\rm LO}$ & $1$ & $-2.2\%$ &  $-2.8\%$ &  $+0.4\%$  &   $+0.6\%$  \\
    \hline
    \end{tabular}
  }
  \caption{\label{tab:fid-cc} $pp \to \mu^+ \nu_\mu +X$.} 
    \end{subtable}
  \caption{\label{tab:fid}
        The different perturbative contributions to the fiducial cross section and their breakdown into the various partonic channels (for $\mu_{\rm R}=\mu_{\rm F}=m_Z$). The numerical uncertainties are stated in brackets.}
\renewcommand{\arraystretch}{1.0}
  \end{table}
%%==========================
The contributions from the channels \mbox{$qg+{\bar q}g$} and
\mbox{$q\gamma+{\bar q}\gamma+g\gamma$} are labelled by $qg$ and
$q(g)\gamma$, respectively.  The contribution from all the remaining
quark--quark channels $qq',\, {\bar q}{\bar q}'$ (including both
\mbox{$q=q'$} and \mbox{$q\neq q'$}) and $q{\bar q}'$ (with
\mbox{$q\neq q'$}) is labelled by $q({\bar q})q^\prime$.  Finally, the
contributions from the gluon--gluon and photon--photon channels are
denoted by $gg$ and $\gamma\gamma$, respectively.

We see that the radiative corrections are subject
to large cancellations between the various partonic channels. The NLO
QCD corrections{ $\sigma^{(1,0)}$} and the NLO EW corrections{
  $\sigma^{(0,1)}$} have a similar size and opposite sign, leading to
an additional cancellation.  The NNLO QCD corrections give an
essentially vanishing contribution within the numerical
uncertainties. The QCD--EW corrections are relatively large and
amounts to $+0.5\%$ with respect to the LO result. 

%%====================================
\begin{figure}[t]
\begin{center}
\includegraphics[width=0.495\textwidth]{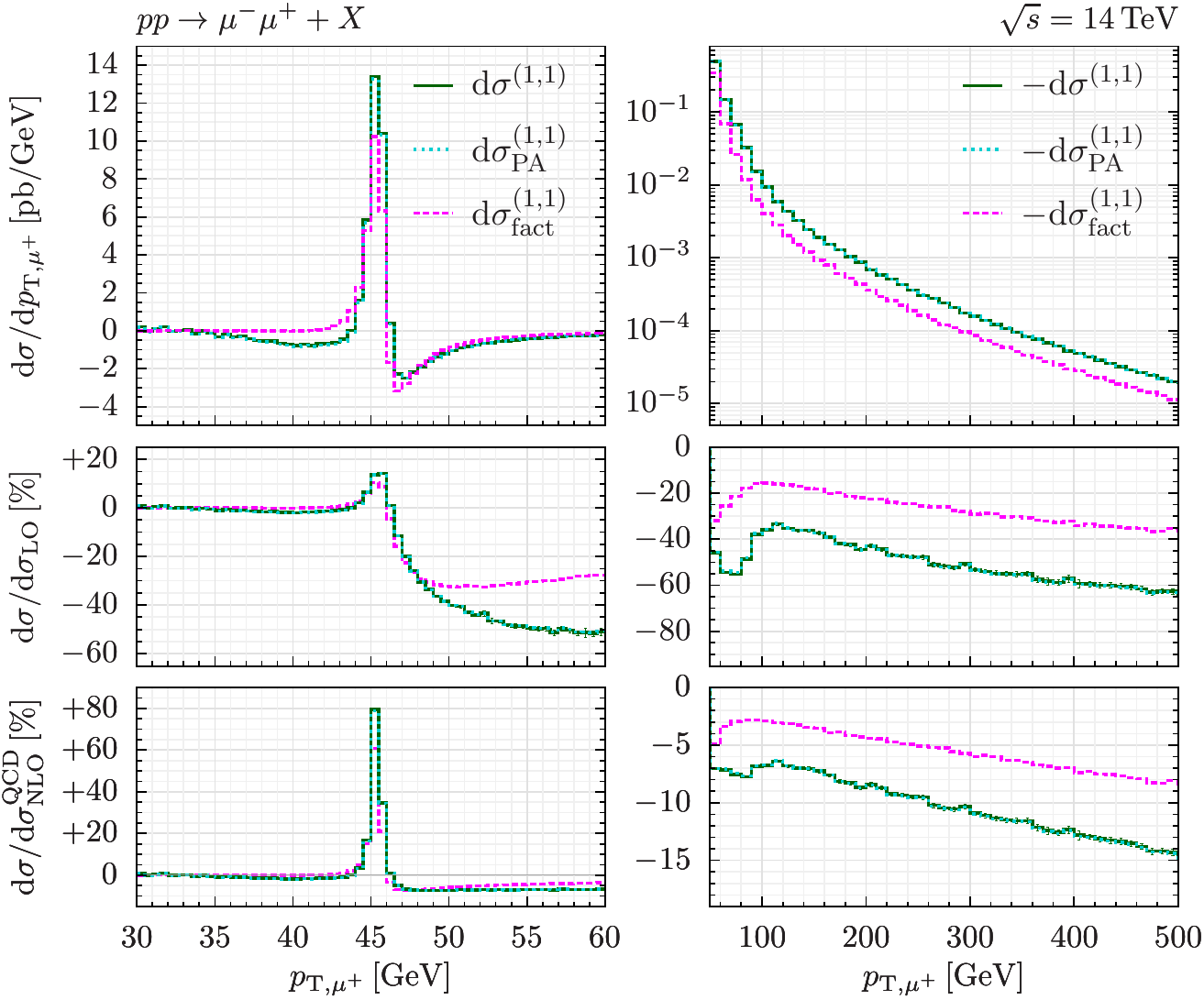} \hfill
\includegraphics[width=0.495\textwidth]{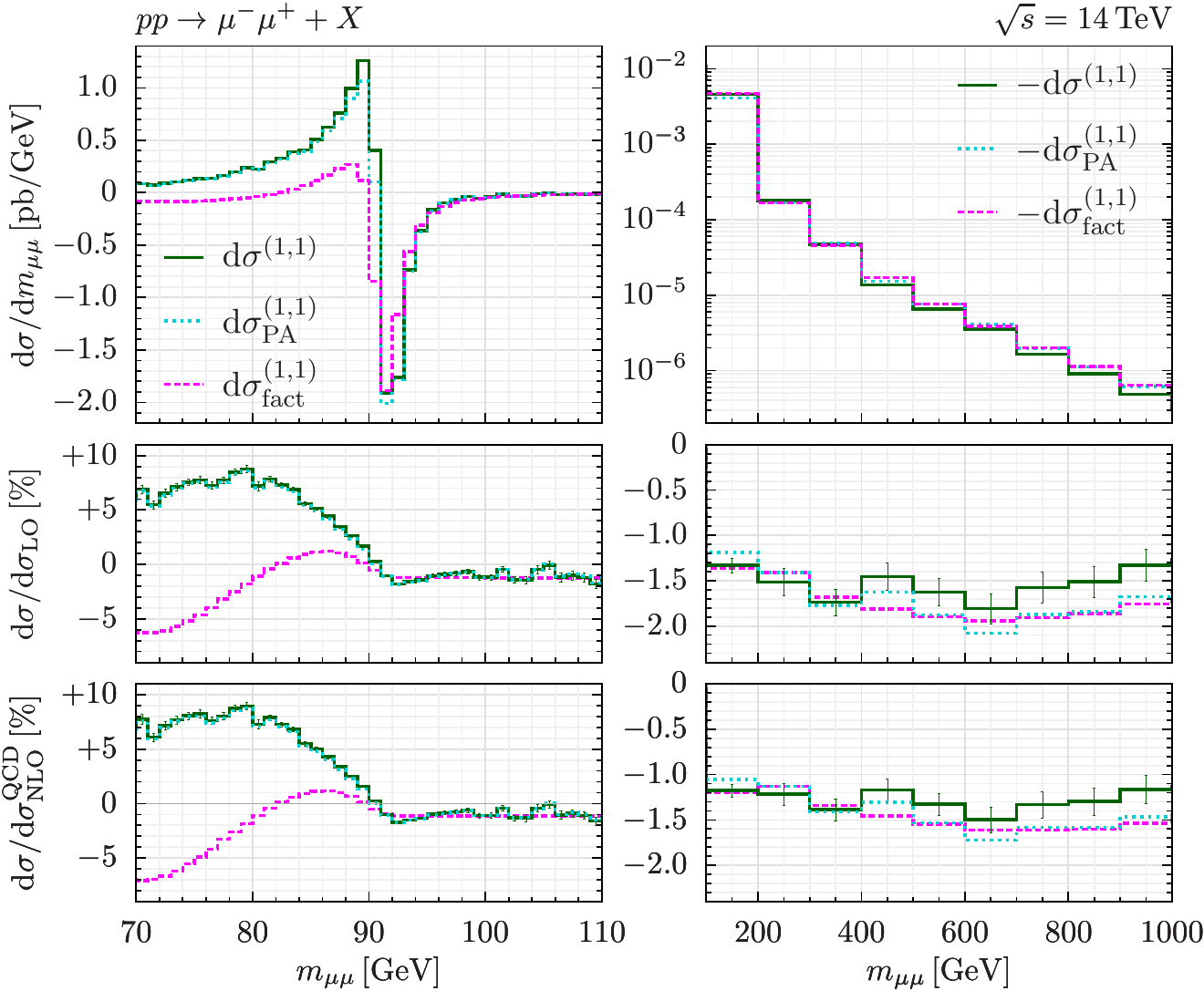}\\[2ex]
\end{center}
\vspace{-2ex}
\caption{\label{fig:spectra-nc} Complete ${\cal O}(\as\alpha)$ corrections to
  the differential cross section $d\sigma^{(1,1)}$ in the anti-muon
  $p_T$ (left) and muon-anti-muon invariant mass $m_{\mu^+\mu^-}$
  (right), compared to the corresponding result in the pole
  approximation and to the factorised approximation
  $d\sigma^{(1,1)}_{\rm fact}$.  The top panels show the absolute
  predictions, while the central (bottom) panels display the ${\cal
    O}(\as\alpha)$ correction normalized to the LO (NLO QCD) result.
  For the full result, the ratios also display the combination of
  statistical and systematic uncertainties associated to the
  subtraction method as explained in Ref.~\cite{Bonciani:2021zzf}. }
\end{figure}
%%====================================
In Fig.~\ref{fig:spectra-nc} we show on the left the ${\cal O}(\as\alpha)$
correction as a function of the anti-muon $p_T$.  The results for the
complete ${\cal O}(\as\alpha)$ correction are compared with those
obtained in two approximations.  The first approximation consists in
computing the finite part of the two-loop virtual amplitude in the
pole approximation~\cite{Dittmaier:2014qza}.  Following the same
strategy adopted for the charged-current Drell-Yan process in
Ref.~\cite{Buonocore:2021rxx}, for which the two-loop amplitude is not
available yet, a suitable reweighting procedure with the exact squared
Born amplitude is performed.

The second approximation is based on a fully factorised approach for
QCD and EW corrections, where we exclude photon-induced processes
throughout (see Ref.~\cite{Dittmaier:2015rxo,Buonocore:2021rxx} for a
detailed description).  We see that the result obtained in the pole
approximation is in perfect agreement with the exact result. This is
due to the small contribution of the two-loop virtual to the computed
correction.

As $p_T$ increases, the (negative) impact of the mixed QCD--EW
corrections increases, and at \mbox{$p_T=500$\,GeV} it reaches about
$-60\%$ with respect to the LO prediction and $-15\%$ with respect to
the NLO QCD result.

In Fig.~\ref{fig:spectra-nc}, we show on the right the result for the
${\cal O}(\as\alpha)$ correction as a function of the di-muon
invariant mass $m_{\mu\mu}$. We notice that the pure factorised result
fails to describe the exact radiative correction below the $Z$
resonance. In contrast, the pole approximation is a very good
approximation of the complete correction, with some small differences
that can be appreciated right around the peak. In the
high-$m_{\mu\mu}$ region the correction is uniformly of the order of
$-1.5\%$ with respect to the NLO QCD result. Here the trend of the
negative correction is captured by both approximations, which,
however, both undershoot the exact result by about $30\%$,
highlighting the relevance of the exact two-loop contribution for this
observable.
\section{Charged current Drell-Yan process}
The mixed QCD--EW corrections to the charged current Drell-Yan process
has been computed in Ref.~\cite{Buonocore:2021rxx}.  All real and
virtual contributions are consistently included but for the two-loop
virtual amplitude. As already mentioned in the previous section, the
latter is approximated by its expansion around the resonant pole,
applying the Pole Approximation and improved via a reweighting
prescription. The quality of this approximation has been discussed in
Ref.~\cite{Buonocore:2021rxx} and strongly confirmed by the comparison
with the exact result for the neutral current case (see previous
section).
\subsection{Results}
\label{sec:results}
We show results for the process $pp\to \mu^+\nu_\mu^-+X$ at centre-of-mass
energy \mbox{$\sqrt{s}=14$\,TeV} with the setup of
Ref.~\cite{Buonocore:2021rxx}. In particular, we use the following selection cuts,
\begin{equation}
  \label{eq:cuts-2}
  p_{T,\mu}>25\,{\rm GeV}\,,\qquad |y_\mu|<2.5\,,\qquad p_{T,\nu}>25\,{\rm GeV}\,,
\end{equation}
and work at the level of {\it bare muons}. In Table~\ref{tab:fid-cc}
we present predictions for the corresponding fiducial cross section at
central scales $\mu_{\rm R}=\mu_{\rm F}=m_Z$. We see large
cancellations among the partonic channels, similarly to what observed
for the neutral current case, see Table~\ref{tab:fid-nc}. Also here,
the mixed QCD--EW corrections are relatively large, being of similar
size to the NNLO QCD ones. Because of the large cancellations
occurring for central scales, we observe that the pattern of the
higher-order QCD corrections to the perturbative series has a strong
dependence on the choice of the renormalization and factorisation
scales.  For example, the scale choice $\mu_R=\mu_F={m_W}/{2}$ leads
to a more common perturbative pattern: $\sigma^{(1,0)}/\sigma_{\rm LO}
= +10\%$, $\sigma^{(0,1)}/\sigma_{\rm LO} = -2.9\%$,
$\sigma^{(2,0)}/\sigma_{\rm LO} =+4.2\%$, $\sigma^{(1,1)}/\sigma_{\rm
  LO} = +0.76\%$.

\begin{figure}[t]
\begin{center}
\includegraphics[width=0.25\textwidth]{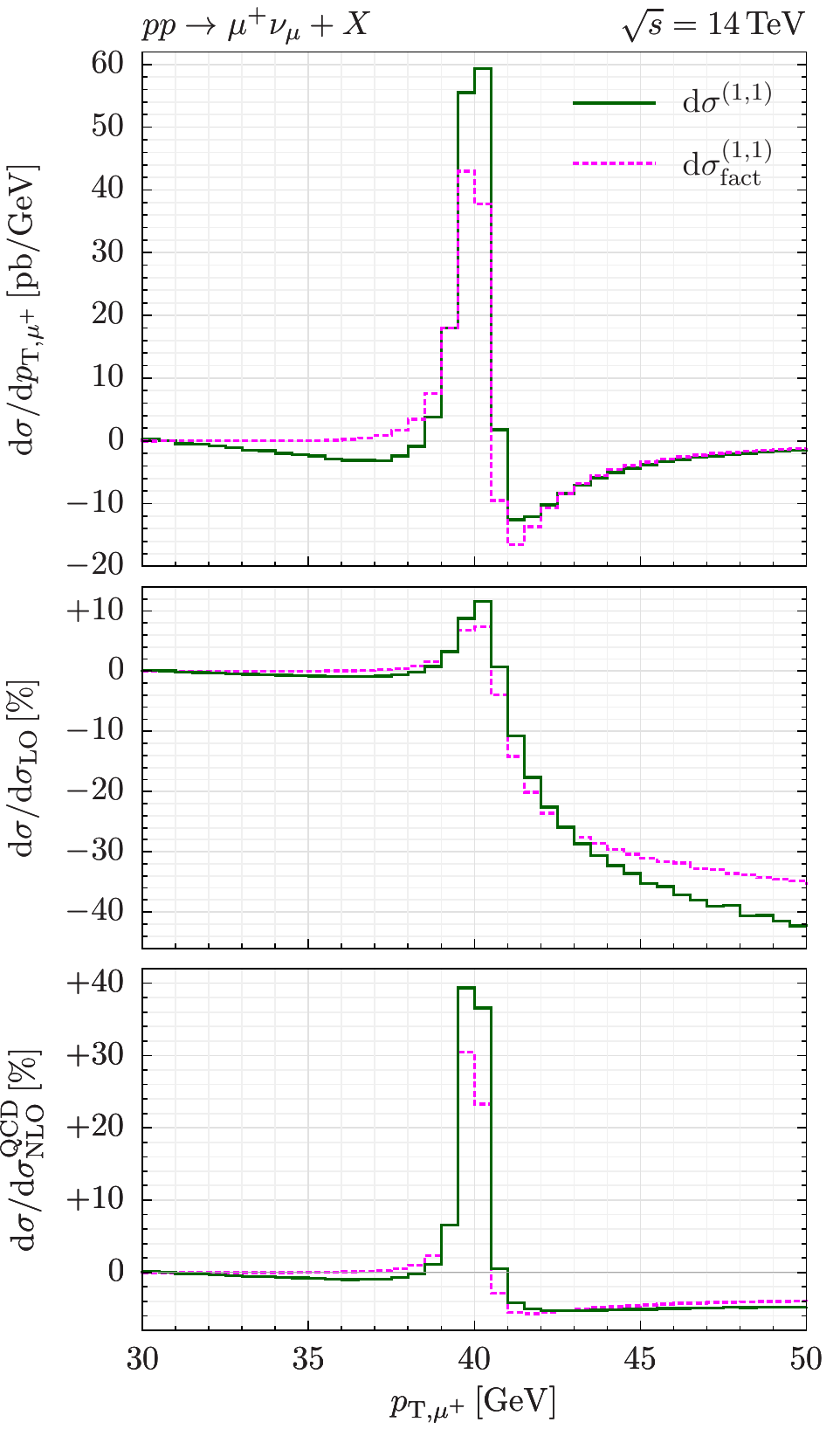}
\includegraphics[width=0.25\textwidth]{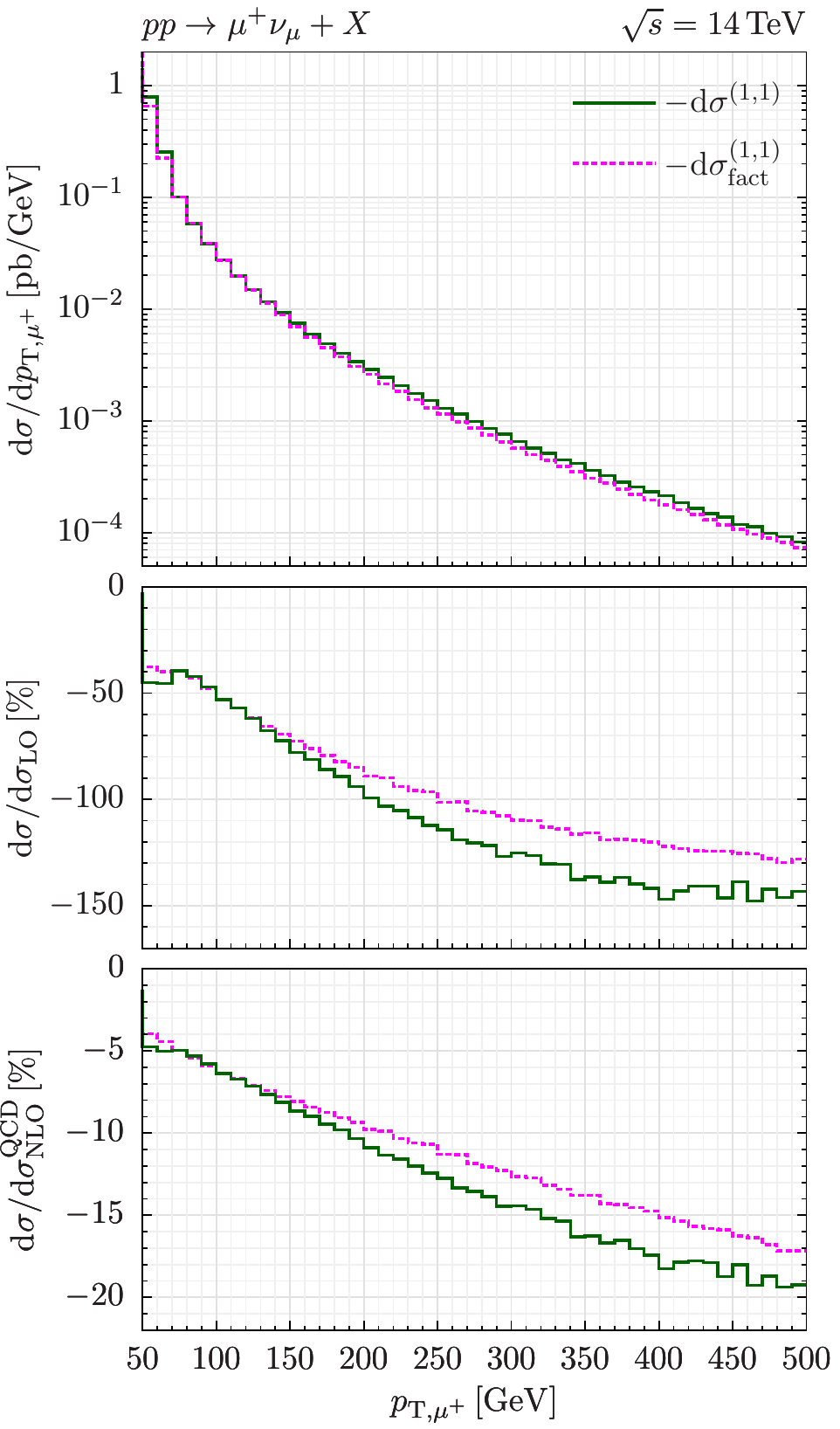}\\
\end{center}
\vspace{-2ex}
\caption{\label{fig:ptOS-OFFS}
  As is in Fig.~\ref{fig:spectra-nc} for the transverse momentum spectrum of the positively charged muon. 
  %% Complete ${\cal O}(\as\alpha)$ correction to the differential cross section $d\sigma^{(1,1)}$ in the muon $p_T$, and its factorized approximation $d\sigma^{(1,1)}_{\rm fact}$, defined in Eq.~(\ref{eq:fact}).
%% The top panels show the absolute predictions, while the central (bottom) panels display the ${\cal O}(\as\alpha)$ correction normalized to the LO (NLO QCD) result.
}
\end{figure}

In Fig.~\ref{fig:ptOS-OFFS}, we show our results for the complete
${\cal O}(\as\alpha)$ correction to the transverse momentum spectrum
of the positively charged muon.  Similarly to the neutral current
case, we compare our results with those obtained with a multiplicative
combination of the NLO QCD and NLO EW corrections.  We observe that
overall the factorised approximation reproduces qualitatively well the
complete results. As $p_T$ increases, the negative impact of the
mixed QCD--EW corrections increases and becomes rather sizeable,
reaching at $p_T=500$\,GeV about $-140\%$ with respect to the LO
prediction and $-20\%$ with respect to the NLO QCD result. This is not
unexpected, since the high-$p_T$ region is dominated by $W+{\rm jet}$
topologies, for which the ${\cal O}(\as\alpha)$ effects can be seen as
NLO EW corrections.
%%====================================
%%====================================

\section{Conclusions}
We have presented results for mixed QCD-EW corrections to both the
neutral current- and the charged current- Drell-Yan process. We have shown
their impact on fiducial cross sections and selected differential
observables, finding relatively large effects. In particular, for the
neutral current case, we have shown the first result at this order that is valid in
the entire range of dilepton invariant masses.

\section*{Acknowledgments}
 This material is based upon work supported in part by the Swiss
 National Science Foundation under contracts IZSAZ2$\_$173357 and
 200020$\_$188464, by the ERC Starting Grant 714788 REINVENT, by the
 INFN, by the Italian \textit{Ministero dell'Universit\`a e della
   Ricerca} Grant No. PRIN2017, by the European Research Council under
 the European Unions Horizon 2020 research and innovation Programme
 Grant Agreement No. 740006 and by the COST Action CA16201
 PARTICLEFACE.

\bibliographystyle{JHEP}
\bibliography{biblio.bib}

\providecommand{\href}[2]{#2}\begingroup\raggedright\begin{thebibliography}{10}

\bibitem{Duhr:2020seh}
C.~Duhr, F.~Dulat and B.~Mistlberger, \emph{{Drell-Yan Cross Section to Third
  Order in the Strong Coupling Constant}},
  \href{http://dx.doi.org/10.1103/PhysRevLett.125.172001}{\emph{Phys. Rev.
  Lett.} {\bfseries 125} (2020) 172001},
  [\href{https://arxiv.org/abs/2001.07717}{{\ttfamily 2001.07717}}].

\bibitem{Chen:2021vtu}
X.~Chen, T.~Gehrmann, N.~Glover, A.~Huss, T.-Z. Yang and H.~X. Zhu,
  \emph{{Di-lepton Rapidity Distribution in Drell-Yan Production to Third Order
  in QCD}},  \href{https://arxiv.org/abs/2107.09085}{{\ttfamily 2107.09085}}.

\bibitem{Duhr:2020sdp}
C.~Duhr, F.~Dulat and B.~Mistlberger, \emph{{Charged current Drell-Yan
  production at N$^{3}$LO}},
  \href{http://dx.doi.org/10.1007/JHEP11(2020)143}{\emph{JHEP} {\bfseries 11}
  (2020) 143}, [\href{https://arxiv.org/abs/2007.13313}{{\ttfamily
  2007.13313}}].

\bibitem{Bonciani:2021zzf}
R.~Bonciani, L.~Buonocore, M.~Grazzini, S.~Kallweit, N.~Rana, F.~Tramontano and
  A.~Vicini, \emph{{Mixed strong$-$electroweak corrections to the Drell$-$Yan
  process}},  \href{https://arxiv.org/abs/2106.11953}{{\ttfamily 2106.11953}}.

\bibitem{Catani:2007vq}
S.~Catani and M.~Grazzini, \emph{{An NNLO subtraction formalism in hadron
  collisions and its application to Higgs boson production at the LHC}},
  \href{http://dx.doi.org/10.1103/PhysRevLett.98.222002}{\emph{Phys. Rev.
  Lett.} {\bfseries 98} (2007) 222002},
  [\href{https://arxiv.org/abs/hep-ph/0703012}{{\ttfamily hep-ph/0703012}}].

\bibitem{Buonocore:2019puv}
L.~Buonocore, M.~Grazzini and F.~Tramontano, \emph{{The $q_T$ subtraction
  method: electroweak corrections and power suppressed contributions}},
  \href{http://dx.doi.org/10.1140/epjc/s10052-020-7815-z}{\emph{Eur. Phys. J.
  C} {\bfseries 80} (2020) 254},
  [\href{https://arxiv.org/abs/1911.10166}{{\ttfamily 1911.10166}}].

\bibitem{Buonocore:2021rxx}
L.~Buonocore, M.~Grazzini, S.~Kallweit, C.~Savoini and F.~Tramontano,
  \emph{{Mixed QCD-EW corrections to $\boldsymbol{pp\!\to\!\ell\nu_\ell\!+\!X}$
  at the LHC}},
  \href{http://dx.doi.org/10.1103/PhysRevD.103.114012}{\emph{Phys. Rev. D}
  {\bfseries 103} (2021) 114012},
  [\href{https://arxiv.org/abs/2102.12539}{{\ttfamily 2102.12539}}].

\bibitem{Catani:2019iny}
S.~Catani, S.~Devoto, M.~Grazzini, S.~Kallweit, J.~Mazzitelli and H.~Sargsyan,
  \emph{{Top-quark pair hadroproduction at next-to-next-to-leading order in
  QCD}}, \href{http://dx.doi.org/10.1103/PhysRevD.99.051501}{\emph{Phys. Rev.
  D} {\bfseries 99} (2019) 051501},
  [\href{https://arxiv.org/abs/1901.04005}{{\ttfamily 1901.04005}}].

\bibitem{Dittmaier:2014qza}
S.~Dittmaier, A.~Huss and C.~Schwinn, \emph{{Mixed QCD-electroweak
  $\mathcal{O}(\alpha_s\alpha)$ corrections to Drell-Yan processes in the
  resonance region: pole approximation and non-factorizable corrections}},
  \href{http://dx.doi.org/10.1016/j.nuclphysb.2014.05.027}{\emph{Nucl. Phys. B}
  {\bfseries 885} (2014) 318--372},
  [\href{https://arxiv.org/abs/1403.3216}{{\ttfamily 1403.3216}}].

\bibitem{Dittmaier:2015rxo}
S.~Dittmaier, A.~Huss and C.~Schwinn, \emph{{Dominant mixed QCD-electroweak
  O($\alpha$$_s$$\alpha$) corrections to Drell\textendash{}Yan processes in the
  resonance region}},
  \href{http://dx.doi.org/10.1016/j.nuclphysb.2016.01.006}{\emph{Nucl. Phys. B}
  {\bfseries 904} (2016) 216--252},
  [\href{https://arxiv.org/abs/1511.08016}{{\ttfamily 1511.08016}}].

\end{thebibliography}\endgroup

%% \begin{thebibliography}{99}
%% \bibitem{...}
%% ....

%% \end{thebibliography}

\end{document}